\documentclass[aps,prd,twocolumn,groupedaddress,floatfix,nofootinbib,showpacs]{revtex4}
\usepackage{amsmath}
\usepackage{amsfonts}
\usepackage{amssymb}
\usepackage{latexsym}
\usepackage{graphicx}
\usepackage{bm}
\usepackage{epsfig}
\usepackage[english]{babel}

\begin{document}


\title{Cosmic Steps in Modeling Dark Energy}
\author{Tower Wang\footnote{Electronic address: wangtao218@pku.edu.cn}}
\affiliation{Center for High-Energy Physics, Peking University,\\
Beijing 100871, China\\ \vspace{0.2cm}}
\date{\today\\ \vspace{1cm}}
\begin{abstract}
The past and recent data analyses gave some hints of steps in dark
energy. Considering the dark energy as a dynamical scalar field, we
investigate several models with various steps: a step in the scalar
potential, a step in the kinetic term, a step in the energy density
and a step in the equation-of-state parameter $w$. These toy models
provide a workable mechanism to generate steps and features of dark
energy. Remarkably, a single real scalar can cross $w=-1$
dynamically with a step in the kinetic term.
\end{abstract}


\maketitle



\section{Introduction}
Dark energy is a popular explanation for the recent acceleration of
our Universe. In order to draw a portrait of the dark energy, it is
necessary to parameterize it and constrain the parameters from
observational data. By the past and recent data analyses, it was
indicated that the dark energy could have various steps in its
parameters. Several years ago, a late-time transition in the
equation of state (EOS) was studied by Bassett \emph{et al}
\cite{Bassett:2002qu}. Very recently, Huang \emph{et al}
\cite{Huang:2009rf} reported that the dark energy might spring out
at a low redshift. However, there were few theoretical studies to
explain such steps in the EOS or in the density of dark energy.

Our purpose here is to make a step toward this direction. We will
treat the dark energy as a dynamical scalar field, and then study a
step in the scalar potential, in the kinetic term, in the energy
density and in the EOS respectively. Lacking of enough observational
data, hitherto the portrait of dark energy is still vague and the
hints of steps are weak, so we do not attempt to build a realistic
model in this paper. Instead, to give a vivid picture of the
mechanism, we will play with simplified toy models and choose
exaggerated model parameters. Because these models are difficult to
solve analytically, for the most time we will rely on numerical
algorithms. The models and algorithms can be easily extended and
refined to give a more realistic description of the dark energy.

Although this could be the first time to systematically study cosmic
steps in dark energy models with explicit Lagrangian, the stepped
model is not novel in cosmology. It is an old story: possibly steps
of the inflaton potential have left some fingerprints at the born
time of our Universe \cite{Adams:2001vc}. Since the dark energy is
more elusive, the physical origin of the stepped dark energy field
is less clear than the stepped inflaton field.

\section{Methodology}
Before going to specific models, we will describe the general
framework and numerical methods in some detail. Impatient readers
can skip directly to the next section to get our models
(\ref{stepV}), (\ref{stepf}), (\ref{potential}) and main results
depicted in figures.

In the absence of spatial fluctuations, the Lagrangian density of a
scalar field minimally coupled to gravity has the form
\begin{equation}\label{lagrangian}
\mathcal{L}_{\phi}=a^3\left[\frac{1}{2}f(\phi)\dot{\phi}^2-V(\phi)\right],
\end{equation}
where $a$ is the scale factor. Here the function $f(\phi)$ in the
kinetic term is new. It is positive for quintessence
\cite{Zlatev:1998tr} and negative for phantom
\cite{Caldwell:1999ew}. Later on we will also discuss a new model in
which $f(\phi)$ evolves dynamically from $+1$ to $-1$. In that
situation, the single real scalar plays the role of quintom
\cite{Feng:2004ad}.

In a flat universe dominated by dark energy together with cold dark
matter, if we ignore the contribution of ordinary matter for
simplicity, then the evolution dynamics is governed by the following
Friedmann equations
\begin{eqnarray}
\nonumber H^2&=&\frac{8\pi G_N}{3}(\rho_{m}+\rho)\\
\label{Friedmann1} &=&\frac{8\pi G_N}{3}\left[\frac{\rho_{m0}a_0^3}{a^3}+\frac{1}{2}f(\phi)\dot{\phi}^2+V(\phi)\right],\\
\nonumber \dot{H}&=&-4\pi G_N(\rho_{m}+\rho+p)\\
\label{Friedmann2} &=&-4\pi G_N\left[\frac{\rho_{m0}a_0^3}{a^3}+f(\phi)\dot{\phi}^2\right].
\end{eqnarray}
Here $\rho_{m}$ is the energy density of dark matter, taking value
$\rho_{m0}$ at the present time with the scale factor $a=a_0$. The
dark energy $\phi$ has an energy density $\rho$ and a pressure $p$,
whose subscripts have been left out for briefness. One should not
mistake $\rho$ and $p$ as the total energy density and the total
pressure. The EOS parameter is defined by $w=p/\rho$ as usual.

In the above we have used a dot to denote the derivative with
respect to comoving (physical) time $t$. For instance, we have taken
the convention of notation $\dot{\phi}=d\phi/dt$ and defined
$H=\dot{a}/a$. It will be convenient to employ the notations
$x=\ln(a/a_0)=-\ln(1+z)$ and $\phi'=d\phi/dx$, then they give us a
useful relation $\dot{\phi}=H\phi'=-H(1+z)\phi_{,z}$. Utilizing
(\ref{Friedmann1}), we find the kinetic energy is
\begin{equation}\label{kinetic}
\frac{1}{2}f\dot{\phi}^2=\frac{8\pi G_Nf\left[\rho_{m0}(1+z)^3+V\right](1+z)^2\phi_{,z}^2}{6-8\pi G_Nf(1+z)^2\phi_{,z}^2}.
\end{equation}

The equation of motion
\begin{equation}\label{eomt}
f\ddot{\phi}+3Hf\dot{\phi}+\frac{1}{2}f_{,\phi}\dot{\phi}^2+V_{,\phi}=0
\end{equation}
for the scalar field can be written as
\begin{eqnarray}\label{eomz}
\nonumber &&4\pi G_Nf(1+z)\left[\rho_{m0}(1+z)^3+4V\right]\phi_{,z}\\
\nonumber &=&8\pi G_Nf(1+z)^2\left[\rho_{m0}(1+z)^3+V\right]\phi_{,zz}\\
\nonumber &&+16\pi^2G_N^2f^2(1+z)^3\left[\rho_{m0}(1+z)^3+2V\right]\phi_{,z}^3\\
\nonumber &&+4\pi G_Nf_{,\phi}(1+z)^2\left[\rho_{m0}(1+z)^3+V\right]\phi_{,z}^2\\
&&+V_{,\phi}\left[3-4\pi G_Nf(1+z)^2\phi_{,z}^2\right].
\end{eqnarray}
Given explicit form of $f(\phi)$ and $V(\phi)$, equation
(\ref{eomz}) alone dictates the evolution of dark energy. But this
is a highly nonlinear second order differential equation. In most
cases we have to resort to numerical methods. We will take two
slightly different schemes, dubbed the linear iteration method and
the cubic iteration method, since they reform equation (\ref{eomz})
into linear algebraic equations or cubic algebraic equations
respectively. Let us elaborate on them now.

For numerically evolving equation (\ref{eomz}) from an initial point
$z_{i}$ to the final point $z_{f}$, we partition the interval
$[z_{i},z_{f}]$ into $N$ equal subintervals of width
$h=z_{j}-z_{j-1}=(z_{f}-z_{i})/N$. It is convenient to notate
$\phi_{j}=\phi(z_{j})$ for short, where $j=0,1,2,...,N$. For
numerical computation, the derivatives can be approximated by finite
differences
\begin{eqnarray}
\label{difference1}\left.\frac{d\phi}{dz}\right|_{z=z_{j}}&=&\frac{\phi_{j+1}-\phi_{j-1}}{2h},\\
\label{difference2}\left.\frac{d^2\phi}{dz^2}\right|_{z=z_{j}}&=&\frac{\phi_{j+2}-2\phi_{j}+\phi_{j-2}}{4h^2}.
\end{eqnarray}
Making use of approximations (\ref{difference1}) and
(\ref{difference2}), one can recast equation (\ref{eomz}) into a
linear algebraic equation with respect to $\phi_{j+2}$. Starting
with the initial values of $\phi_{j}$ at $j=0,1,2,3$, we can get the
values of all $\phi_{j}$ with $j>3$ iteratively. The linear
iteration method gives a unique path of $\phi$ for numerical
evolutions. Its disadvantage is the excessive number of initial
conditions. Remember that for a second order differential equation
we usually impose two initial conditions. In practical operation, we
treat with potentials flat at the initial point $\phi_{i}$, and set
$\phi_{0}=\phi_{1}=\phi_{2}=\phi_{3}=\phi_{i}$.

Rather than (\ref{difference2}) one may tend to estimate the second
order derivative with
\begin{equation}\label{difference3}
\left.\frac{d^2\phi}{dz^2}\right|_{z=z_{j}}=\frac{\phi_{j+1}-2\phi_{j}+\phi_{j-1}}{h^2}.
\end{equation}
Now we need only two initial conditions, for instance,
$\phi_{0}=\phi_{1}=\phi_{i}$. This is the starting point of the
cubic iteration scheme. In such a scheme, instead of a linear
equation of $\phi_{j+2}$, one has to solve a cubic equation with
respect to $\phi_{j+1}$. The formula of roots of a cubic equation is
well known. For our purpose, it is enough to treat them in two
general categories: (\emph{i}) if the equation has one real root and
a pair of imaginary roots, then $\phi_{j+1}$ takes a value of the
real root; (\emph{ii}) if all roots are real, we determine the value
of $\phi_{j+1}$ by minimizing $|\phi_{j+1}-\phi_{j}|$. The trick
enables us to pick out the smoothest evolution path of $\phi$, and
fortunately this path is unique in our simulation. Even if the path
is not unique after applying the above trick, one can still find the
smoothest path by further minimizing
$|\phi_{j+1}+\phi_{j-1}-2\phi_{j}|$, \emph{etc}.

The above two methods are operated independently in our algorithms.
As double checks, they agree with each other very well. In fact, the
resulting graphs look like duplicates. During the numerical
simulation, we define the reduced Planck mass $M_{pl}=1/\sqrt{8\pi
G_N}$, the fractional density of dark energy
$\Omega_{\mathrm{DE}}=\rho_{\mathrm{DE}}/(3M_p^2H^2)$ and the
present fractional density of dark matter
$\Omega_{m0}=\rho_{m0}/(3M_p^2H_0^2)$, where $H_0$ is Hubble
parameter at the present time. Moreover, we set $\Omega_{m0}=0.3$
and work in the unit $M_{pl}=1$. In a flat universe, ignoring the
contribution of ordinary matter, one has
$\Omega_{\mathrm{DE}}+\Omega_{m}=1$. For every specific model below,
we evolve equation (\ref{eomz}) from $z_{i}=20$ to $z_{f}=0$ with
$N=2000$ subintervals. Fixing the initial conditions and other
parameters, if we increase either $N$ or $z_{i}$, the change in
results is unobservable. This confirms the reliability of our
methods above and results below.

\section{Models with Steps}
\begin{figure}
\begin{center}
\includegraphics[width=0.45\textwidth]{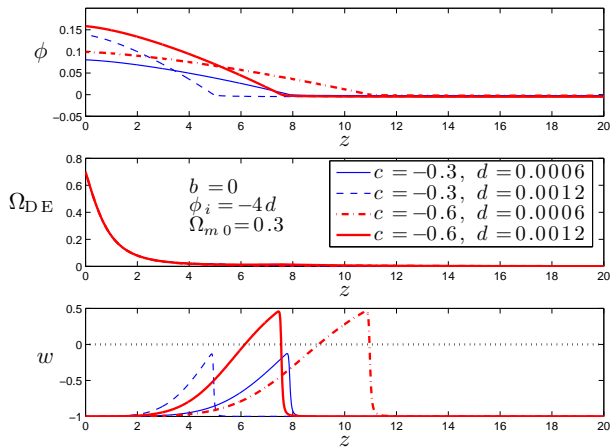}\\
\end{center}
\caption{(color online). Evolution curves of the quintessence field
with potential (\ref{stepV}) and $f=+1$. There are bumps in the
curves of fractional density and EOS parameter, although the bump in
fractional density is almost unnoticeable. The amplitude and
location of bumps can be tuned by changing parameters in our model.
We set $8\pi G_N=1$ in all figures.}\label{fig-quintessence}
\end{figure}
\begin{figure}
\begin{center}
\includegraphics[width=0.45\textwidth]{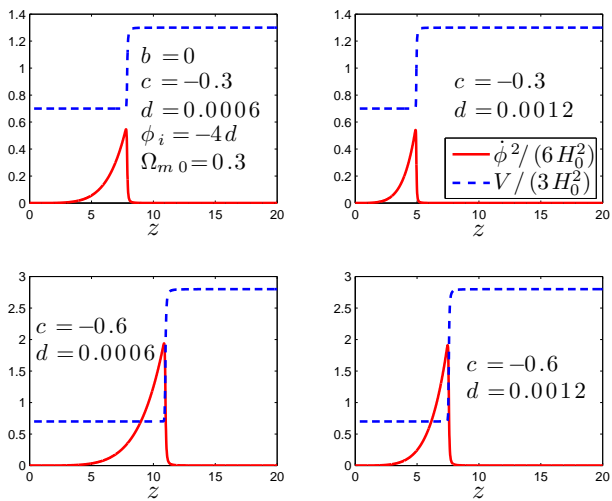}\\
\end{center}
\caption{(color online). The rise and fall of the kinetic energy and
potential of a stepped quintessence with potential (\ref{stepV}) and
$f=+1$. Both the kinetic energy and the potential are normalized by
$3H_0^2$, where $H_0$ is the Hubble parameter at
$z=0$.}\label{fig-kinV}
\end{figure}
\begin{figure}
\begin{center}
\includegraphics[width=0.45\textwidth]{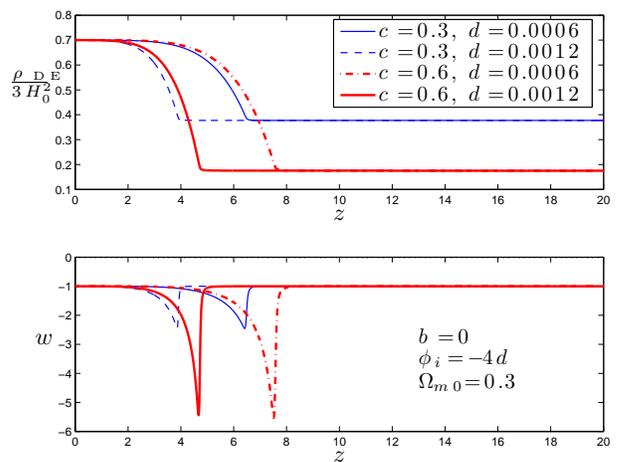}\\
\end{center}
\caption{(color online). For stepped phantom with potential
(\ref{stepV}) and $f=-1$. The dark energy density climbs up a step,
while EOS parameter has a dip nearby. Similarly to the stepped
quintessence, the amplitude and location of the steps and dips are
tunable as we change the model parameters.}\label{fig-rhow}
\end{figure}
\begin{figure}
\begin{center}
\includegraphics[width=0.45\textwidth]{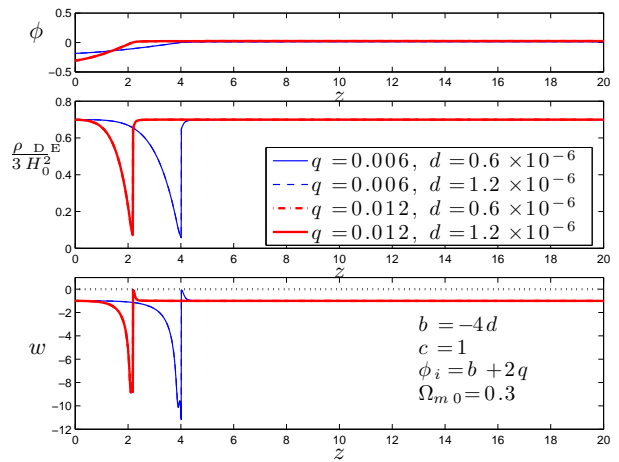}\\
\end{center}
\caption{(color online). The evolution of the scalar field, dark
energy density and EOS parameter for the stepped quintom model
(\ref{stepf}). In contrast with previous figures, a small change in
parameter $d$ does not modify the evolution curves
significantly.}\label{fig-quintom}
\end{figure}

\subsection{Steps in the Scalar Potential}
So far we have not specified our models. It will be done in this
section. Firstly, suppose the potential of dark energy is broadly
flat but has a sudden transition near $\phi=b$. Then it is natural
to model such a stepped potential in the following form
\begin{equation}\label{stepV}
V(\phi)=V_0\left[1+c\tanh\left(\frac{\phi-b}{d}\right)\right].
\end{equation}
For relatively small $d$, the hyperbolic tangent function is a good
smooth analytic approximation to a step function. The parameter $b$
determines the location of step in field space, while $c$ and $d$
control the height and width of step respectively. Based on this
model, more complicated ones can be obtained by superposing the
steps or replacing the constant $V_0$ with other functions. Note
here $c$ is a dimensionless parameter, but $b$ and $d$ are in unit
of reduced Planck mass $M_{pl}=1/\sqrt{8\pi G_N}$. However, in our
simulation and figures, we will simply set $M_{pl}=1$.

We study this potential in the context of quintessence with $f=+1$
and phantom with $f=-1$ respectively. Since
$\Omega_{\mathrm{DE}0}+\Omega_{m0}=1$ in a flat universe with
negligible ordinary matter, it is easy to prove
$V_0=3H_0^2(1-\Omega_{m0})/(1\pm|c|)$ for quintessence (lower sign)
and phantom (upper sign). The simulation results for stepped
quintessence are shown in figures \ref{fig-quintessence} and
\ref{fig-kinV}. The results for phantom are depicted in figure
\ref{fig-rhow}. In plotting them we have chosen the step location
$b=0$, and the initial value of $\phi$ is chosen as $\phi_{i}=-4d$
at redshift $z_i=20$. From the figures we can see a step in the
energy density. Quintessence falls down the step but phantom climbs
up. Although the location of step in field space is fixed by $b=0$,
its location in redshift space is dependent of $c$ and $d$. At the
same location there is a feature (a bump for quintessence or a dip
for phantom) in curves of kinetic energy and EOS parameter.
Amplitudes of steps and features are determined mainly by parameter
$c$. The initial value $\phi_{i}$ of the scalar field resides well
in the flat region of the potential because $\tanh(-4)\simeq-0.999$.
If we increase the absolute value of $\phi_{i}$, the steps and
features will shift to lower redshift region.

From figures \ref{fig-kinV} and \ref{fig-rhow}, one can roughly read
the location $z_T$ and width $\Delta_T$ of the step as well as the
initial density $\rho_i$ and final density $\rho_f$ of the dark
energy. In data analysis, it is useful to capture these properties
by parameterizing the dark energy density as
\begin{eqnarray}\label{steprho}
\nonumber \rho&=&\rho_i+\frac{\rho_f-\rho_i}{1+\exp\left(\frac{z-z_T}{\Delta_T}\right)}\\
&=&\frac{\rho_i+\rho_f}{2}+\frac{\rho_i-\rho_f}{2}\tanh\left(\frac{z-z_T}{2\Delta_T}\right).
\end{eqnarray}
This parametrization is applicable when the dark energy changes from
$\rho_i$ to $\rho_f$ near the redshift $z_T$. Assuming the dark
energy is decoupled with other ingredients, we have the continuity
equation $\dot{\rho}+3H(\rho+p)=0$ or equivalently
$(\ln\rho)'=-3(1+w)$. Corresponding to (\ref{steprho}), it gives
\begin{equation}
w=\frac{(1+z)(\rho_i-\rho_f)\exp\left(\frac{z-z_T}{\Delta_T}\right)}{3\Delta_T\left[1+\exp\left(\frac{z-z_T}{\Delta_T}\right)\right]\left[\rho_f+\rho_i\exp\left(\frac{z-z_T}{\Delta_T}\right)\right]}-1.
\end{equation}
One may check that the function gives rise to a bump if
$\rho_i>\rho_f$ or a dip when $\rho_i<\rho_f$, in agreement with
figures \ref{fig-quintessence} and \ref{fig-rhow}.

\subsection{Steps in the Kinetic Term}
Having discussed the steps in the dark energy potential, we discuss
what will happen if there is a step in the coefficient $f(\phi)$ of
the kinetic term. It is interesting to study a new model
\begin{eqnarray}\label{stepf}
\nonumber f(\phi)&=&c\tanh\left(\frac{\phi-b}{d}\right),\\
V(\phi)&=&V_0\left[1+\tanh\left(\frac{\phi^2}{q^2}\right)\right].
\end{eqnarray}
In favor of the fact that $\Omega_{\mathrm{DE}0}+\Omega_{m0}=1$, we
can set $V_0=3H_0^2(1-\Omega_{m0})/2$. In the potential $V(\phi)$
there is a deep dip, whose depth has been fixed. The coefficient
$f(\phi)$ is non-trivial. It has a step located at $\phi=b$ with
amplitude $c$ and width $d$. Again $c$ is a dimensionless parameter,
while $b$ and $d$ are in unit of reduced Planck mass $M_{pl}$.  As
mentioned before, we set $M_{pl}=1$ in our simulation and figures.

The idea is to settle the potential bottom $\phi=0$ in the region
$f>0$ by ensuring $c\tanh(-b/d)>0$. Given an appropriate initial
condition, the scalar first rolls down to the bottom of potential
and then moves up, getting closer and closer to the point $\phi=b$.
However, this model still requires some fine-tuning, because it is
not always possible for the scalar field to pass the sign-inversion
point of $f(\phi)$ after leaving the bottom of potential. We
fine-tune the parameters, and then numerically evolve the equation
of motion (\ref{eomz}). According to the simulation results in
figure \ref{fig-quintom}, both the density and the EOS parameter of
dark energy decrease near the sign-inversion point. In particular,
the EOS parameter jumps down abruptly from $w>-1$ to $w\ll-1$. This
is a simple way to cross the EOS barrier $w=-1$. To make it we need
only one real scalar field, completely relying on its own dynamics.
However, one should not regard (\ref{stepf}) as more than a toy
model. There is an infamous instability in any model with a
wrong-signed kinetic term, \emph{e.g.}, in the phantom model
\cite{Caldwell:1999ew}. We suspect the stepped quintom model here
suffers from the same problem. Various issues on this class of model
were explored in \cite{Vikman:2004dc,Nojiri:2005pu,Xia:2007km}.

\subsection{Steps in the EOS Parameter}
If there is a step in the EOS parameter, one usually parameterizes
$w$ in a form \cite{Bassett:2002qu} similar to (\ref{steprho}).
However, for building a model later, we parameterize it in an
alternative way,
\begin{equation}\label{stepw}
w=w_i+\frac{w_f-w_i}{1+\left(\frac{1+z}{1+z_T}\right)^\lambda}=w_i+\frac{w_f-w_i}{1+\left(\frac{a_T}{a}\right)^{\lambda}},
\end{equation}
where $\lambda\gg1$. With this form, the continuity equation can be
integrated out as
\begin{eqnarray}
\nonumber \rho&=&\rho_T\left(\frac{a_T}{a}\right)^{3(1+w_i)}\left[\frac{1}{2}+\frac{1}{2}\left(\frac{a}{a_T}\right)^{\lambda}\right]^{3(w_i-w_f)/\lambda}\\
&=&\rho_T\left(\frac{a_T}{a}\right)^{3(1+w_f)}\left[\frac{1}{2}+\frac{1}{2}\left(\frac{a_T}{a}\right)^{\lambda}\right]^{3(w_i-w_f)/\lambda}.
\end{eqnarray}

Note there is a duality $w_i\leftrightarrow w_f$,
$\lambda\leftrightarrow-\lambda$ in (\ref{stepw}). We are most
interested in the following special cases:
\begin{itemize}
\item $w_i=-1$; Now the EOS takes the form
\begin{equation}\label{eosi}
\frac{p}{\rho}=w_f-\frac{1+w_f}{2}\left(\frac{\rho}{\rho_T}\right)^{\lambda/[3(1+w_f)]}.
\end{equation}
\item $w_f=-1$. In this case, we reexpress the EOS as
\begin{equation}\label{eosf}
\frac{p}{\rho}=w_i-\frac{1+w_i}{2}\left(\frac{\rho_T}{\rho}\right)^{\lambda/[3(1+w_i)]}.
\end{equation}
\end{itemize}
The equations of state (\ref{eosi}) and (\ref{eosf}) can be casted
into a unified form
\begin{equation}\label{pressure}
p=\rho(w_c-A\rho^{\alpha}),
\end{equation}
from which one can integrate the continuity equation to give
\begin{equation}\label{rho}
\rho^{\alpha}=\frac{1+w_c}{A+Ba^{3\alpha(1+w_c)}}=\frac{1+w_c}{A\left[1+\left(\frac{a}{a_T}\right)^{3\alpha(1+w_c)}\right]}.
\end{equation}

\begin{figure}
\begin{center}
\includegraphics[width=0.35\textwidth]{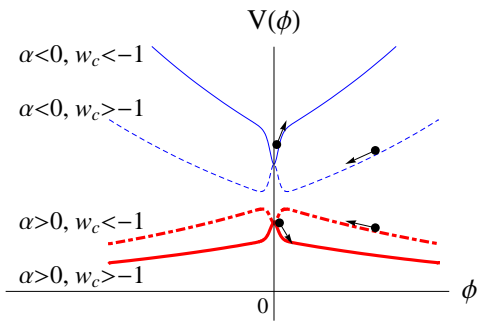}\\
\end{center}
\caption{(color online). A rough illustration of potential
(\ref{potential}) with different parameter choices. The arrows
indicate evolving directions of the dark energy field. For example,
when $\alpha<0$ and $w_c<-1$, the potential is illustrated by the
thin solid blue line, and the scalar field rolls up the potential
from the bottom $\phi=0$ to $\phi\rightarrow\pm\infty$, while the
EOS parameter $w$ evolves from $-1$ to $w_c$ asymptotically. The
dashed blue line corresponds to the potential with $\alpha<0$ and
$w_c>-1$, in which case the scalar field rolls down the potential
from $\phi\rightarrow\pm\infty$ to one of the local minimum point,
and the EOS parameter $w$ evolves form $w_c$ to
$-1$.}\label{fig-potential}
\end{figure}

It is time for us to reconstruct the potential of (\ref{lagrangian})
from (\ref{stepw}) in a universe dominated by dark energy. To do
this, we assume $f(\phi)$ is a nonzero constant and $\rho_m=0$,
$w_c\neq-1$. For details on reconstructing the potential of scalar
dark energy, please refer to
\cite{Saini:1999ba,Sahni:2006pa,Capozziello:2005tf,Elizalde:2008yf}.
Making use of equations (\ref{Friedmann1}), (\ref{Friedmann2}),
(\ref{pressure}) and (\ref{rho}), we finally obtain
\begin{eqnarray}\label{potential}
&&a^{3\alpha(1+w_c)}=\frac{A}{B}\sinh^2\left(\frac{\phi}{\phi_c}\right),\\
\nonumber &&V(\phi)=\frac{V_c}{2}\cosh^{-2/\alpha}\left(\frac{\phi}{\phi_c}\right)\left[2-(1+w_c)\tanh^2\left(\frac{\phi}{\phi_c}\right)\right]
\end{eqnarray}
with $\phi_c^{-1}=\alpha\sqrt{6\pi G_Nf(1+w_c)}$ and
$V_c=[(1+w_c)/A]^{1/\alpha}$. For various choices of parameters, the
shape of this potential is illustrated in figure
\ref{fig-potential}. The EOS parameter $w\rightarrow w_c$
asymptotically as $\phi\rightarrow\pm\infty$. Near the bottom or top
of the potential, it approaches $-1$. The field rolls down the
potential as a quintessence if $w_c>-1$, but rolls up as a phantom
when $w_c<-1$. We impose $|3\alpha(1+w_c)|\gg1$ by hand to
accomplish the sudden EOS transition (\ref{stepw}).

\section{Discussion}
Treating dark energy as a scalar field, we explicitly modeled steps
in its potential, kinetic term, density and EOS, and thus provided a
workable mechanism to explain the previously and recently claimed
dark energy transitions. To arrive at a realistic model, more
experimental and theoretical efforts are needed in the future. When
this work was near completion, a related paper
\cite{Mortonson:2009qq} appeared, where they arrived at models with
an implicit potential similar to the shape of $V(\phi)$
in(\ref{stepf}). As a partial list, some previous papers relevant to
our work are
\cite{Zhou:2007xp,Capozziello:2005mj,Nojiri:2006ww,Nojiri:2006jy,Nojiri:2005sx,Szydlowski:2006qn,Hrycyna:2009zj}.

\begin{acknowledgments}
This work is supported by the China Postdoctoral Science Foundation.
We are grateful to Tong Li, Wei Liao for discussions and Qinyan Tan
for help in programme.
\end{acknowledgments}

\end{document}